# MASTER OT J190519.41+301524.4:
# New Eclipsing Cataclysmic Variable of VY Scl Type

F. Martinelli[1], D. V. Denisenko[2*]

[1] Astronomical Center Lajatico, Italy
[2] Sternberg Astronomical Institute, Lomonosov Moscow State University, Russia

**Abstract**–MASTER OT J190519.41+301524.4 was discovered as an optical transient of 15.7$^m$ by the Mobile Astronomical System of TElescope-Robots in March 2014. We report the results of photometric observations of this variable performed at Lajatico Astronomical Center in June-July 2015. The light curve is showing deep V-shaped eclipses with an amplitude of two magnitudes. The orbital period was determined to be 0.129694 d (3.113 h). Based on the archival observations and the shape of the orbital curve we suggest that MASTER OT J190519.41+301524.4 is a new cataclysmic variable of VY Scl type ("anti-nova") with an inclination angle close to 90 deg.

## INTRODUCTION

MASTER OT J190519.41+301524.4 is an optical transient in Lyra discovered on Mar. 06, 2014 (Denisenko et al., 2014) by MASTER-Kislovodsk auto-detection system (Lipunov et al., 2010). The previously unremarkable star was found to be highly variable on the digitized Palomar plates. Namely, the star was bright on 1950 June 17 POSS-I plate, but very faint on 1987 June 21 POSS-II plate (see Figure 1). Those changes of brightness are clearly seen from the magnitudes reported in various catalogs. Namely, USNO-B1.0 catalogue (Monet et al., 2003) is listing USNO-B1.0 1202-0321874 with $B1$=16.32 $R1$=15.65 $B2$=17.84 $R2$=N/A $I$=16.39, while GSC 2.3.2 is giving Fmag=19.41 measured from 1987 June 21 Palomar plate. The star is an ultraviolet source GALEX J190519.4+301525 with the far and near UV magnitudes FUV=17.66±0.05, NUV=17.46±0.04. Based on this information Denisenko et al. suggested that MASTER OT J190519.41+301524.4 is a cataclysmic variable, most likely the anti-nova of VY Scl type in the high state.

In this article we report the results of our observations of MASTER OT J190519.41+301524.4 obtained in June-July 2015 at Lajatico Astronomical Center in Italy. We have detected the deep eclipses in this compact binary system, measured its orbital period and confirmed the classification as a VY Scl cataclysmic variable.

## PHOTOMETRY

The observations of J1905+3015 were performed at Lajatico Astronomical Center in Italy http://www.astronomicalcentre.org/ on seven nights (2015 June 28, 30, July 7, 12, 15, 17 and 18) using 0.36-m Cassegrain telescope with SBIG ST-8XME CCD. A total of 203 unfiltered images with 300-sec exposures were obtained (26, 11, 35, 23, 34, 35 and 37 images covering 2.9, 1.4, 3.7, 2.8, 4.1, 3.0 and 3.8 hours, respectively). Nearby star USNO-B1.0 1202-0321923 (0.6' East and 0.2' South of the variable) with the magnitude V=12.8 was used as a reference star.

* E-mail: d.v.denisenko@gmail.com

The observing times were converted from JD to Barycentric Julian Date using online period search service http://scan.sai.msu.ru/lk/ by Kirill Sokolovsky. Using Lafler-Kinman and Deeming methods we have obtained the best value of period 0.129694(2) d, or 3.113 hr. The phased light curve from our observations is presented in Figure 2. It is showing deep eclipses with a total amplitude about 2 magnitudes and the variations by $0.3^m$ near the maximum light. Search for the secondary periods has not given any additional periodic signals at higher frequencies.

Light elements of J1905+3015 obtained from our observations:

$$\text{Min} = \text{HJD } 2457202.428 + 0.129694 \times E$$

## DISCUSSION

The deep fading of J1905+3014 observed on the 1987 POSS-II Red plate can be explained either by the low state of a polar (magnetic CV without an accretion disk) or by the drop of accretion rate in the variable of VY Scl type ("anti-nova" with the disk). Our observations are telling strongly in favor of the second scenario. The V-shape and long duration of the eclipse (about 20 per cent of the orbital period) are consistent with the presence of quite a large accretion disk. The system was also not detected by ROSAT all-sky survey (Voges et al., 2003), as one would have expected for a magnetic CV in the 16-18$^m$ range.

MASTER OT J190519.41+301524.4 has turned out to be an eclipsing cataclysmic variable of VY Scl type. Systems like that are quite rare. Catalog of cataclysmic binaries by Ritter and Kolb (2003; Version 7.23 of July 2015) lists only eight nova-like variables with eclipses (NL+E), but none of them is classified as "anti-nova". Interestingly enough, all of them have orbital periods longer than 0.133 d. The phased light curve of J1905+3015 binary system reminds that of V482 Cam = HS 0728+6738 (Rodriguez-Gil et al., 2004) that belongs to the subgroup of eclipsing SW Sextantis stars. It is possible that J1905+3015 is also an SW-type variable. We encourage the continued monitoring of this new CV to follow its return to the low state. When the system fades back to 18-19$^m$ level, it will become a good target to measure the size of the accretion disk and to determine the orbit inclination from the changes in eclipse depth and duration.

FIGURES

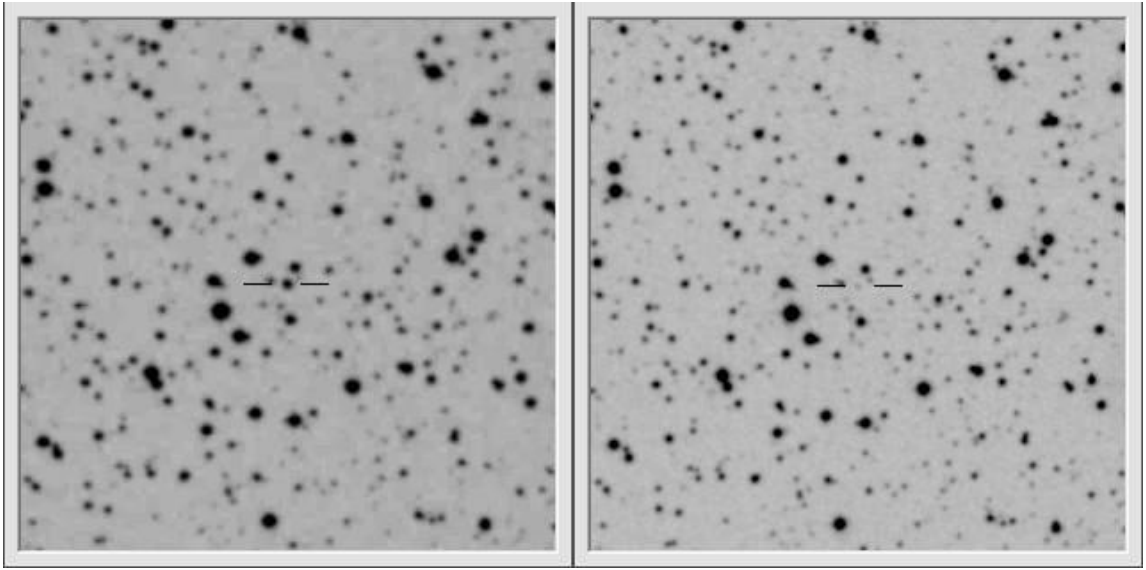

Figure 1. MASTER OT J190519.41+301524.4 on the digitized Palomar Red plates. Left: 1950 June 17 (bright state), right: 1987 June 21 (faint). FOV is 5'x5', North is up and East is at left.

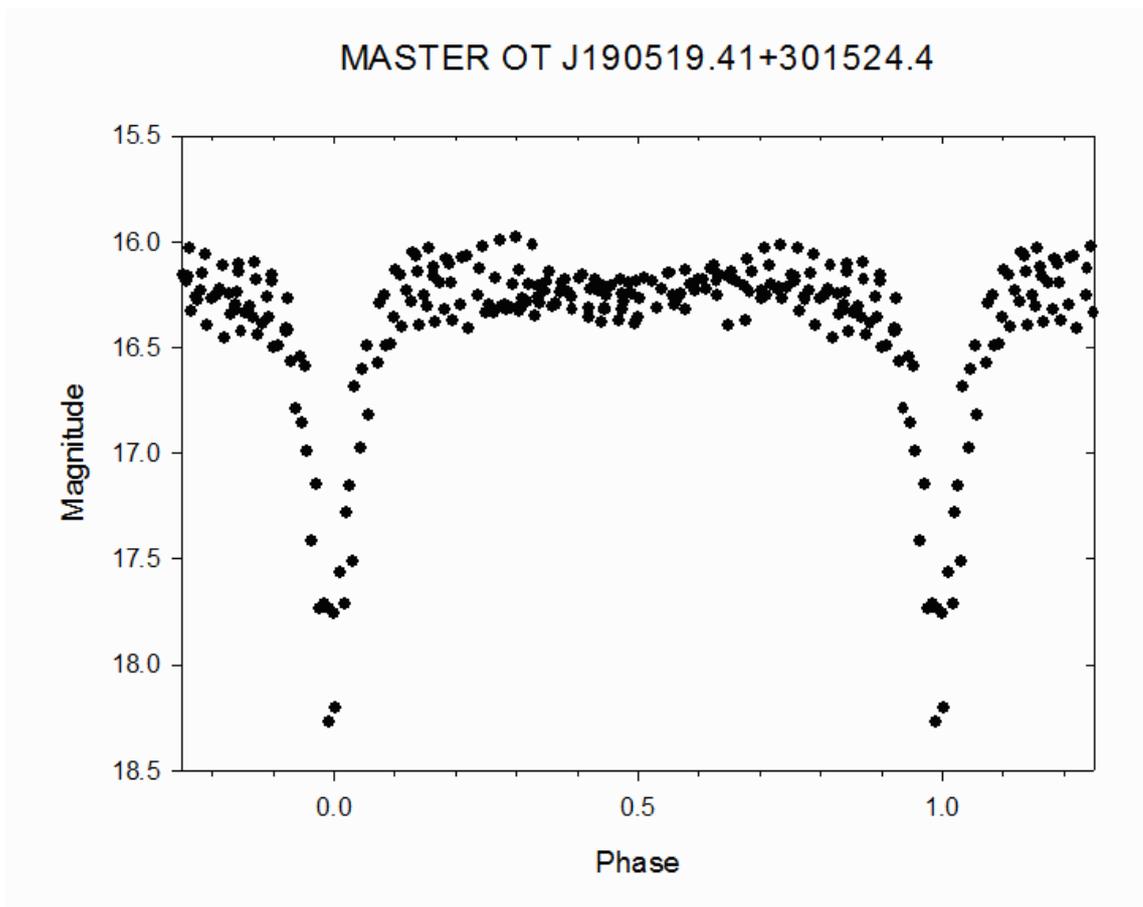

Figure 2. Light curve of J1905+3015 from Lajatico data (June-July 2015) folded with the best orbital period P=0.129694 d and the initial epoch $T_0$= 2457202.428 (HJD).